 [2003/12/12 5.2/AAS markup document class]%

\documentclass{emulateapj}

\shorttitle{Voids in the Galaxy Distribution}
\shortauthors{Thompson and Gregory}

\begin{document}


\title{An Historical View:  The Discovery of Voids in the Galaxy Distribution}


\author{Laird A. Thompson}
\affil{Astronomy Department, University of Illinois Urbana-Champaign, 1002 W. Green
Street, Urbana, IL 61801}
\author{Stephen A. Gregory}
\affil{Physics and Astronomy Department, University of New Mexico, 800 Yale Blvd. NE, Albuquerque, NM 87131, and Boeing LTS Inc., Kirkland AFB, NM 87185}
\email{thompson@astro.illinois.edu, sgregory58@gmail.com}



\begin{abstract}
Voids in the large scale distribution of galaxies were first
recognized and discussed as an astrophysical phenomenon in two
papers published in 1978.  We published the first (Gregory and
Thompson 1978) and Joeveer, Einasto and Tago (1978) published the second.
The discovery of voids altered the accepted view of the large scale
structure of the universe.  In the old picture, the universe
was filled with field galaxies, and occasional density enhancements
could be found at the locations of rich galaxy clusters or
superclusters.  In the new picture, voids are interspersed between
complex filamentary supercluster structure that forms the so-called
cosmic web.  The key observational
prerequisite for the discovery of voids was a wide-angle redshift survey
displayed as a cone diagram that extended far enough in distance to
show a fair sample of the universe (i.e. well beyond the Local
Supercluster).  The initial impact of the 1978 discovery of voids was
stunted for several years by theoretical cosmologists in the West who
were not quite sure how filaments and voids could emerge from what
otherwise appeared to be a homogeneous universe.  After it became clear
several years later that theoretical models of structure formation could
explain the phenomenon, the new void and supercluster paradigm became
widely accepted.

\end{abstract}


%% subject header.  Objects should be in the appropriate "individual"
%% headers (e.g. quasars: individual, stars: individual, etc.) with the
%% additional provision that the total number of headers, including each
%% individual object, not exceed six.  The \objectname{} macro, and its
%% alias \object{}, is used to mark each object.  The macro takes the object
%% name as its primary argument.  This name will appear in the paper
%% and serve as the link's anchor in the electronic edition if the name
%% is recognized by the data centers.  The macro also takes an optional
%% argument in parentheses in cases where the data center identification
%% differs from what is to be printed in the paper.



\section{Introduction}
In the mid-1970's two independent research programs revealed the
beautiful void and supercluster structure of the universe, what is
now referred to as the cosmic web.  We initiated the first of these
research programs (Gregory and Thompson 1978).  The second was an
entirely separate effort by observational astronomers in Estonia and
theoreticians in Russia (Joeveer, Einasto and Tago 1978). These two
programs could not have been more different, but they both led to
the conclusion that there are large volumes in the local universe
with diameters $\sim$20 h$^-$$^1$ Mpc that contain no galaxies
whatsoever\footnote{h = Hubble constant/100 km s$^-$$^1$
Mpc$^-$$^1$}. We called these empty regions ``voids'' while Joeveer
et al. at first simply called them ``big holes''.

Our work was empirically driven. The existence of supercluster
structure was widely known in this era, and Abell (1961) had called
attention to ``second order clusters'', i.e. congregations of rich
cluster cores. Our aim was to understand the galaxy distribution in
and around these collections of Abell clusters. For example, we
asked whether adjacent Abell cluster cores are isolated density
enhancements or whether they are interconnected by ``bridges'' of
galaxies. Because the nearest rich Abell clusters are distributed
within a volume that extends to distances of $\sim$100 h$^-$$^1$
Mpc, we designed our redshift surveys to sample the appropriate
volume. Our first target was the Coma and A1367 cluster pair, and it
was in the large foreground volume between the Milky Way and the
Coma/A1367 supercluster that we identified the first voids. Our work
was not related to nor driven by any theoretical model of structure
formation.

The Estonian work was aimed at testing a specific model of structure
formation championed in the 1970's by the prominent Russian
cosmologist Yakov Zeldovich.  Zeldovich (1970) postulated the
so-called top-down model of galaxy formation. In his model, large
supercluster-sized structures are the first to form, and individual
galaxies fragment out of the larger collapsing structures.  These
models have not stood the test of time, but in the mid-1970's
Joeveer, Einasto and Tago (1978) collaborated with Zeldovich and
mapped the galaxy and cluster distribution (using catalogues
containing known galaxy redshifts) over volumes similar to those we
were studying. Further historical details about this research program
can be found in Einasto (2009).

In the 1970's the hierarchical clustering model (so-called bottom-up
model) was most popular in the West, but in its original form it could not
explain, ab initio, the existence of voids as large as those
reported in Gregory and Thompson (1978).  At first, those who held
fast to the early hierarchical model acknowledged that the observed
``holes'' exist in the galaxy distribution, but they postulated that
they appeared naturally as a result of random statistical processes.
When we presented our evidence for voids as discrete astrophysical
entities with borders defined by filamentary structures, we
encountered skepticism that is documented most clearly in Soneira
and Peebles (1978) where they said the following: "We know that the
eye does tend to judge in a biased way--for example, one readily
picks out ``chains'' of points in a uniform random distribution". In
other words, the filamentary structures that we identified as the
walls of voids were, according to their interpretation, nothing more
than false visual constructs. Not everyone held this extreme view,
of course, but the immediate impact of the early redshift survey
work was diminished despite the fact that the voids that we
identified in the Coma/A1367 foreground (and their filamentary
walls) have stood the test of time and are just as dramatic today as
they were when we first saw them in the mid-1970's.

Fortunately, in this same era advances were being made in computer
simulations aimed at modeling the growth and evolution of structure
in the large scale distribution of galaxies. These models generally
used as a starting point a simple galaxy distribution but
eventually included dark matter (either hot or cold).
Even these early evolutionary models showed how dramatic
filamentary structure can develop in the galaxy distribution over
time, and by doing so, they removed the theoretical prejudice
against the existence of voids. At first the numerical simulations
were relatively simple (Aarseth et al. 1979 and Doroskevich et al.
1980), but the level of sophistication rapidly increased so that by
1983 it was possible for the first time to begin realistically
testing key features of the large scale galaxy distribution (Melott
1983, Frenk et al. 1983, to mention just two).

This brief summary provides a context for understanding how the
discoveries that emerged from the early galaxy redshift surveys
influenced theoretical models of structure formation. Our aim in
what follows is to document the redshift surveys published in the
first decade after this research began.

\section{Early Redshift Surveys}

The redshift survey revolution of the 1970's began when high-voltage
image intensifier devices came into wide use and displaced the
photographic plate as the primary detector in astronomical
spectroscopy.  To measure a redshift accurate to $\pm$100 km/s, a
photographic exposure of $\sim$2.5 hours had previously been
required for an m$\sim$15 galaxy.  An image intensifier system could
produce a similar result in 10 to 15 minutes. Significant numbers of
new redshifts began to be published soon after the Kitt Peak
National Observatory 2.1-m telescope and the Steward Observatory
2.3-m telescope were both equipped with image tube spectrographs. It
is at this point that the Redshift Survey Timeline begins (Table 1).

Key participants in the early work included the late Herb Rood as
well as Guido Chincarini, William Tifft, Stephen Gregory, Laird
Thompson, and Massimo Tarenghi.  Gregory and Thompson had been
graduate students of Tifft at the University of Arizona.  Tarenghi
was Tifft's postdoctoral researcher.  Herb Rood (then at Michigan State
University) and Guido Chincarini (then at the University of Oklahoma) formed a
separate team.

The earliest redshift surveys have been called pencil beams because
target galaxies were selected from cluster cores that span a small
solid angle.  Table 1 contains a complete list of the galaxy
redshift surveys (including the pencil beam surveys) published in the
period 1971-1981.
Hints of the large scale structure first began to
appear as irregularities in the redshift distribution in the pencil
beam foregrounds.  But to see such structure with clarity, a
redshift survey had to probe to m$\sim$15. Surveys to m$\sim$14
showed no significant structure because their distance range was
inadequate, and they could not sample foreground structure in any
detail. With perfect hindsight, any observer who pushed to m$\sim$15
in a pencil beam survey might have said that they saw initial hints
of the structure, but no one immediately grasped its meaning.

A complete description of the large scale structure required
formulating a
new three-dimensional picture of the galaxy distribution, one that
includes voids interspersed between supercluster structure.  The
concept of a galaxy supercluster was well developed by the 1970's
(c.f., Oort 1983), and yet those who accepted the existence of
superclusters assumed that they were simple density
enhancements embedded in a uniform field of background galaxies.
Others suggested that superclusters had a core-halo structure with a
halo that slowly merged into the uniformly distributed field
population of galaxies. So it was the discovery of voids -- and not
superclusters -- that provided the basis for the paradigm change.

\begin{figure}
\epsscale{0.8} \plotone{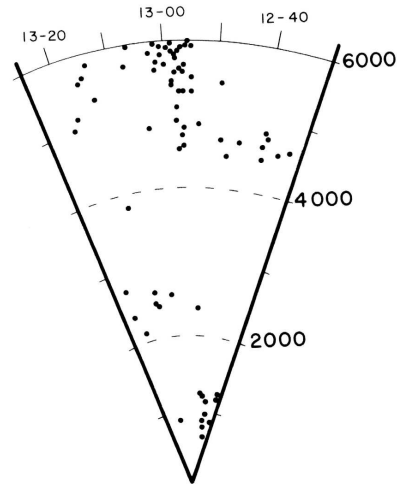} \caption{Tifft and Gregory (1976)
presented this diagram to display their redshift survey for the Coma
cluster to r=6$^o$.  This was the first use of a cone diagram in the
Astrophysical Journal.\label{fig1}}
\end{figure}

The primary catalyst for this discovery was the cone diagram:  a
polar plot with redshift used for the radial coordinate and a
galaxy's angular position on the sky used as the azimuth (the third
dimension being projected into the plane of the plot). Of course, all
modern redshift surveys like CfA, SDSS, 2dF, etc. rely on this diagram to
display their data.  Those who used the cone diagram for the first
time (Tifft and Gregory 1976; Gregory and Thompson 1978; Joeveer,
Einasto and Tago 1978) were the first to visualize the void and
supercluster structure. Table I lists how all authors plotted
their redshift data.

In Chinicarini and Rood's first redshift survey paper, they completed
a pencil beam survey for the Perseus cluster and then
determined its virial mass.  Their second and
third papers list data from the KPNO 2.1-m telescope, and then for
three years they published no redshifts. In the mean time Tifft
began his work at the Steward Observatory 2.3-m telescope.  When
Chincarini and Rood resumed their work with a study of the Coma
cluster, Tifft and graduate student Gregory were doing same. Tifft
and Gregory worked in the central regions of Coma (to r = 6$^o$)
while Chincarini and Rood aimed to trace the Coma cluster to a
radial distance of 14$^o$ by surveying primarily to the west of the
cluster core.  Both groups aimed to collect complete samples to m =
15, and they began the transition away from the narrow pencil beam surveys.

Fig. 1 shows the Tifft and Gregory (1976) Coma cone diagram.  This
is the first redshift survey shown as a cone diagram. In their
discussion Tifft and Gregory (1976) note the lack of field galaxies,
and in the caption to their cone diagram, they say that the
foreground is ``devoid'' of galaxies. The Tifft and Gregory survey
of Coma covered too narrow an angular span to include a complete
void (from one wall through the empty region and on to the other
wall).  An average void spans 20 h$^-$$^1$ Mpc and at the most distant extent
they surveyed 12.6 h$^-$$^1$ Mpc.

Chincarini and Rood published papers in 1975 and 1976 (just before
and just after Tifft and Gregory 1976), but they always plotted
their redshifts as a function of radial distance from the core of
Coma, and their plots were square.  By making a radial analysis,
they essentially destroyed 3D information.  In fact, Chincarini
and Rood never discussed 3D structure in any of the papers they
published prior to 1978. They talked about ``redshift segregation''
whenever their survey intercepted first a void and then a
supercluster filament in the foreground, but they never gave a
physical interpretation as to what it meant, i.e. they never
mentioned concepts like ``holes'' or ``voids''.
Rood and Chincarini discussed their views of the large
scale galaxy distribution most clearly at the end of their last
observational paper on the Coma cluster (Chincarini and Rood 1976a).
They used the following words:  ``The large sizes of clusters and
their fading into low-density supercluster backgrounds leave little
if any space between them.  On the other hand, Figure 3 clearly
shows also a pronounced effect of segregation of redshifts.''

\begin{figure}[b!]
\epsscale{1.0} \plotone{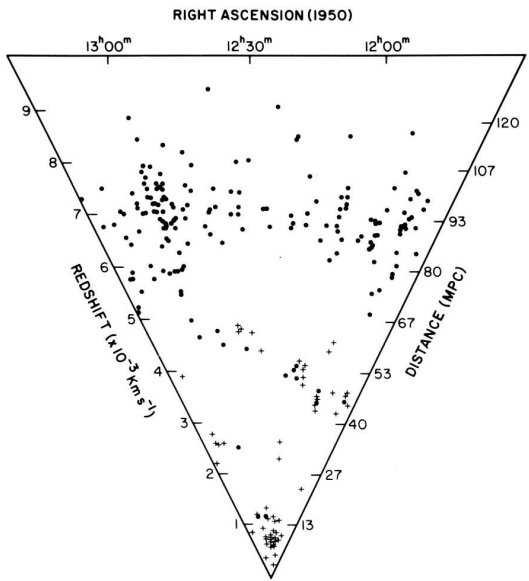} \caption{The Coma/A1367 supercluster
region from Gregory and Thompson (1978). The core of Coma is the
elongated feature on the far left and A1367 the weaker elongated
feature on the far right.  A bridge of galaxies connects the cluster
cores, and in the foreground are the first recognized
voids.\label{fig2}}
\end{figure}

In May, 1975, we submitted
an observing proposal to Kitt Peak National
Observatory to use the 2.1-m telescope for a new redshift survey. In
our proposal, we posed the hypothesis that the two rich Abell
clusters – Coma and A1367 – are enveloped in a common supercluster.
Since they are separated by 21$^o$ in the plane of the sky, our aim
was to survey a slice in the intercluster region between Coma and
A1367 with dimensions 4$^o$ wide and 19$^o$ long to a depth of m =
15.0. We predicted that
we would detect an over-density of galaxies (a bridge at their
common redshift) in the region between the two clusters. This proposal
was accepted, and we collected the new redshifts in April, 1976.

The redshifts in our Coma/A1367 survey were measured and plotted
in a cone diagram by the early summer, 1976 (reproduced here as Fig.
2).  This plot displays the entire galaxy sample in the region that
surrounds and includes Coma and A1367 (11$^o$ x 23$^o$) and not the
smaller area (4$^o$ x 19$^o$) mentioned in our observing proposal.
Fig. 2 is the first wide-angle cone diagram that displays a complete
magnitude-limited sample to m = 15.0.  It is not a pencil beam
survey: at the deepest extent it spans 36 h$^-$$^1$ Mpc across the sky.
Upon seeing this plot we immediately realized the
significance of the irregular distribution of galaxies that had
appeared in the foregrounds of the early pencil beam surveys.

Our Coma/A1367 paper (Gregory and Thompson 1978) arrived at the
Astrophysical Journal on September 7, 1977.  It discusses for the
first time the large scale structure using the new paradigm.  We
list here the key points that are unique to this paper.
\begin{itemize}
\item We recognized huge empty regions in the 3D plot of our survey
volume and for the first time used the word ``void'' to describe them.
\item We outlined possible hypotheses that might
explain the void phenomenon by using the
following words:  ``It is an important challenge for any
cosmological model to explain the origin of these vast, apparently
empty regions of space.  There are two possibilities: (1) the
regions are truly empty, or (2) the mass in these regions is in some
form other than bright galaxies.  In the first case, severe
constraints will be placed on theories of galaxy formation because
it requires a careful (and perhaps impossible) choice of both
$\Omega$ (the present mass density/closure density) and the spectrum
of initial irregularities in order to grow such large density
irregularities...''  It seemed impossible to us at the time
because cold dark matter had not yet been proposed.
\item The abstract to our paper states: ``there are large
regions of space with radii r $>$ 20h$^-$$^1$ Mpc where there appear
to be no galaxies whatever.''
\item Before the observations were made, we had hypothesized
the existence of a bridge of galaxies between Coma
and A1367.  Our 3D cone diagram confirmed it.  Today this bridge of galaxies
is a small segment of what is often called ``The Great Wall''.
\item The general structure shown in our cone diagram includes
the body, the right leg and right arm of what some call the ``Coma stickman''.
\end{itemize}
The sharp contrast between the description of the large
scale distribution of galaxies as given by Gregory and Thompson
(1978) and that of Chincarini and Rood (1976a) explains why a
paradigm change occurred.

Table 1 includes only papers that appeared in refereed journals and
omits papers that were not refereed: observatory publications,
presentations made at meetings like the American Astronomical
Society, and at conferences like the Tallin conference (IAU Symposium
No. 79) held in Tallin, Estonian SSR, September 12-16, 1977\footnote{Neither
Gregory nor Thompson attended the Tallin conference.  Gregory submitted
a request to attend, but his request was denied.}.
The papers presented at this conference were published in
``The Large Scale Structure of the Universe'' (Longair and Einasto
1978). This conference was an important event in the early
study of the large scale distribution of galaxies.  Milikel Joeveer
and Jaan Einasto discussed preliminary
results that they later published in a refereed journal as Joever,
Einasto, and Tago (1978).  Our Coma/A1367 supercluster manuscript arrived
at the Astrophysical Journal five days before the conference began, and we
had no prior knowledge of any presentations that would be made at IAU Symposium
No. 79 except for one by Tifft and Gregory that
briefly mentions Gregory and Thompson (1978).

\begin{figure}
\epsscale{1.0} \plotone{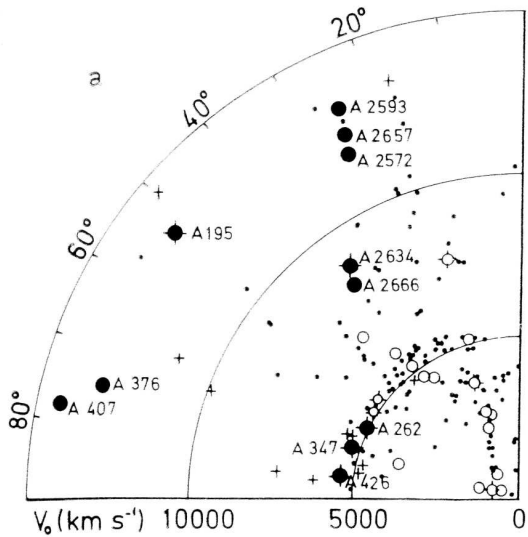} \caption{Redshift cone diagram of
the south galactic hemisphere by Joeveer, Einasto, and Tago (1978).
Large filled circles represent clusters of galaxies.  Small filled
circles represent galaxies. They note the existence of "large holes"
in the 3D distribution.\label{fig3}}
\end{figure}

The Joeveer, Einasto and Tago (1978) paper is an interesting study
of the 3D distribution of both galaxies and clusters of galaxies in
the south galactic hemisphere. They present several cone diagrams
that display the large scale structure in a region of sky
70$^o$x70$^o$. Their cone diagrams show large empty regions, regions
that they call ``big holes''. Because their galaxy redshift data
were taken from previously published sources (the major source being
the Second Reference Catalogue of Bright Galaxies by de Vaucouleurs
et al. 1976), Joeveer et al. (1978) do not claim to present a
complete magnitude-limited survey.
Our surveys always aimed to be magnitude-limited because critics in
the 1970's were quick to raise the possibility that empty
regions appeared empty due to incomplete sampling.
The Joeveer et al. (1978) paper was published
in the November issue of Monthly Notices of the Royal Astronomical
Society, five months after our paper on
Coma/A1367.  Their paper contains four cone
diagrams (four different cuts through the same survey area) as well as
a reference to Gregory and Thompson (1978). Figure
3 below displays one of these four cuts.

We continued our program to study the galaxy distribution in and around
the nearest Abell clusters.  In Gregory, Thompson and Tifft (1981), we investigated the properties of the long filamentary chain of rich clusters in Perseus, and in Gregory and Thompson (1984) we studied another double cluster, A2197 and A2199.  A larger collection of the early redshift survey workers analyzed the properties of the Hercules supercluster region (Tarenghi et al. 1979), and in Chincarini, Thompson and Rood (1981) an extended filament or bridge of galaxies was found to stretch
44h$^-$$^1$ Mpc from the A2197/A2199 supercluster to the Hercules
supercluster.  This publication 
is significant because
it demonstrates the filamentary nature of the large scale structure
over an extensive length scale.  In this case, the filament extends
primarily in depth rather than in an angular span across the sky.  We note that none of these studies from the early 1980's are so-called pencil beam surveys.  The survey areas were large enough to include multiple Abell cluster cores and revealed other voids and supercluster structure.

It is a common misconception that the early phases of the Center for
Astrophysics (CfA) redshift survey contributed to the paradigm
change.  In Table 1 it is easy to see how the CfA work progressed
relative to the work of the Arizona groups.  The first CfA study by
Davis, Geller and Huchra (1978) had a limiting magnitude of m =
13.0. It was shallow and showed no evidence of any structure.
Their stated aim in undertaking this first survey was to measure
the mean mass density of the universe.  The second CfA milestone
paper was by Davis, Huchra, Latham and Tonry (1982).  With a survey
limit at m = 14.5, it shows hints of the large scale structure.  But
by the time this paper was published in 1982, the nature of voids
and supercluster structure was widely known and widely discussed.
For example, Zeldovich, Einasto and Shandarin (1982) published a review
article for Nature entitled ``Giant Voids in the Universe''.
In addition to technical references cited in Table 1, popular
accounts were also being published (Chincarini and Rood 1980;
Gregory and Thompson 1982).  The later article for Scientific
American is entitled ``Superclusters and Voids in the
Distribution of Galaxies''.

\begin{figure*}[t!]
\epsscale{1.0} \plottwo{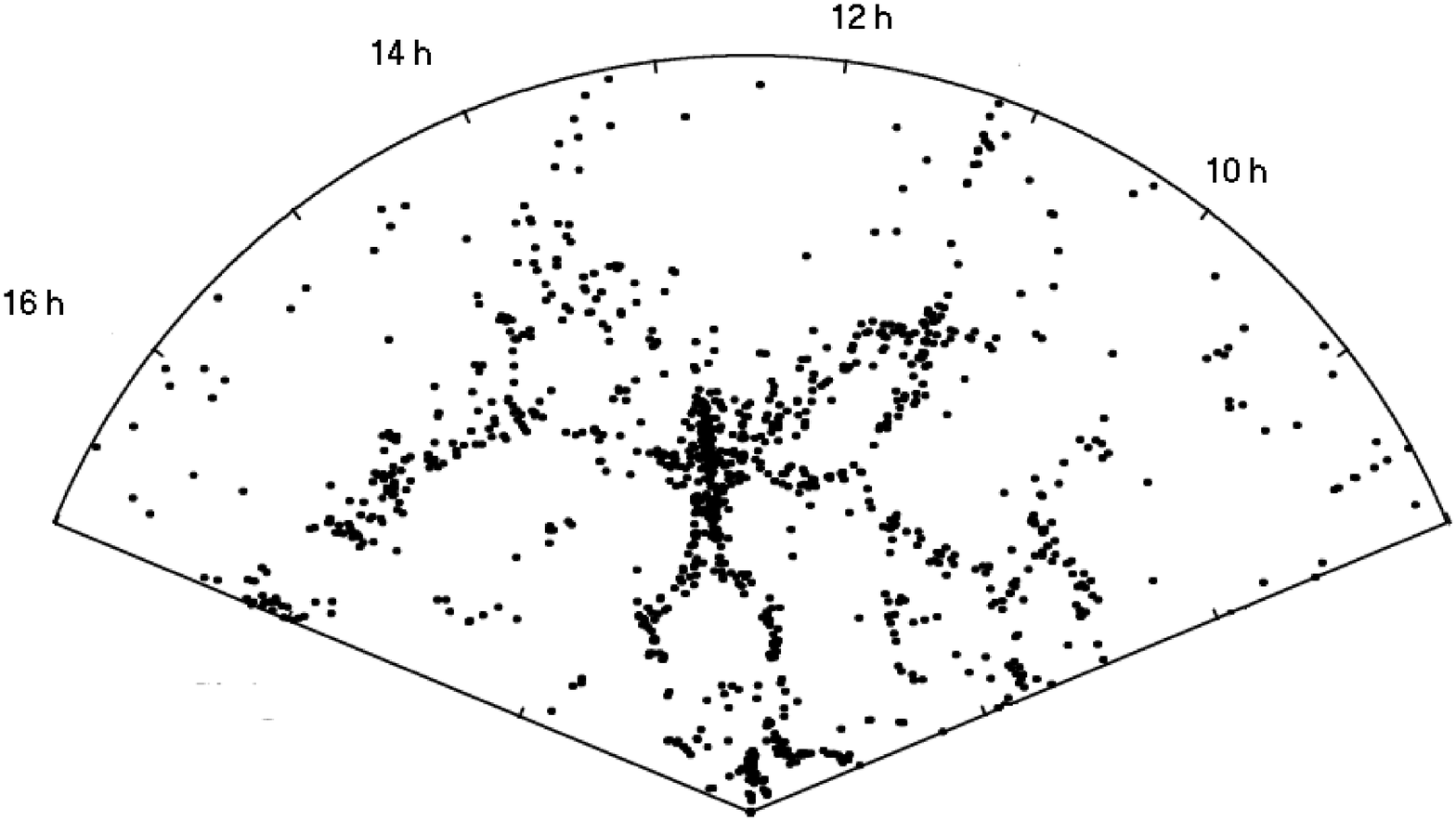}{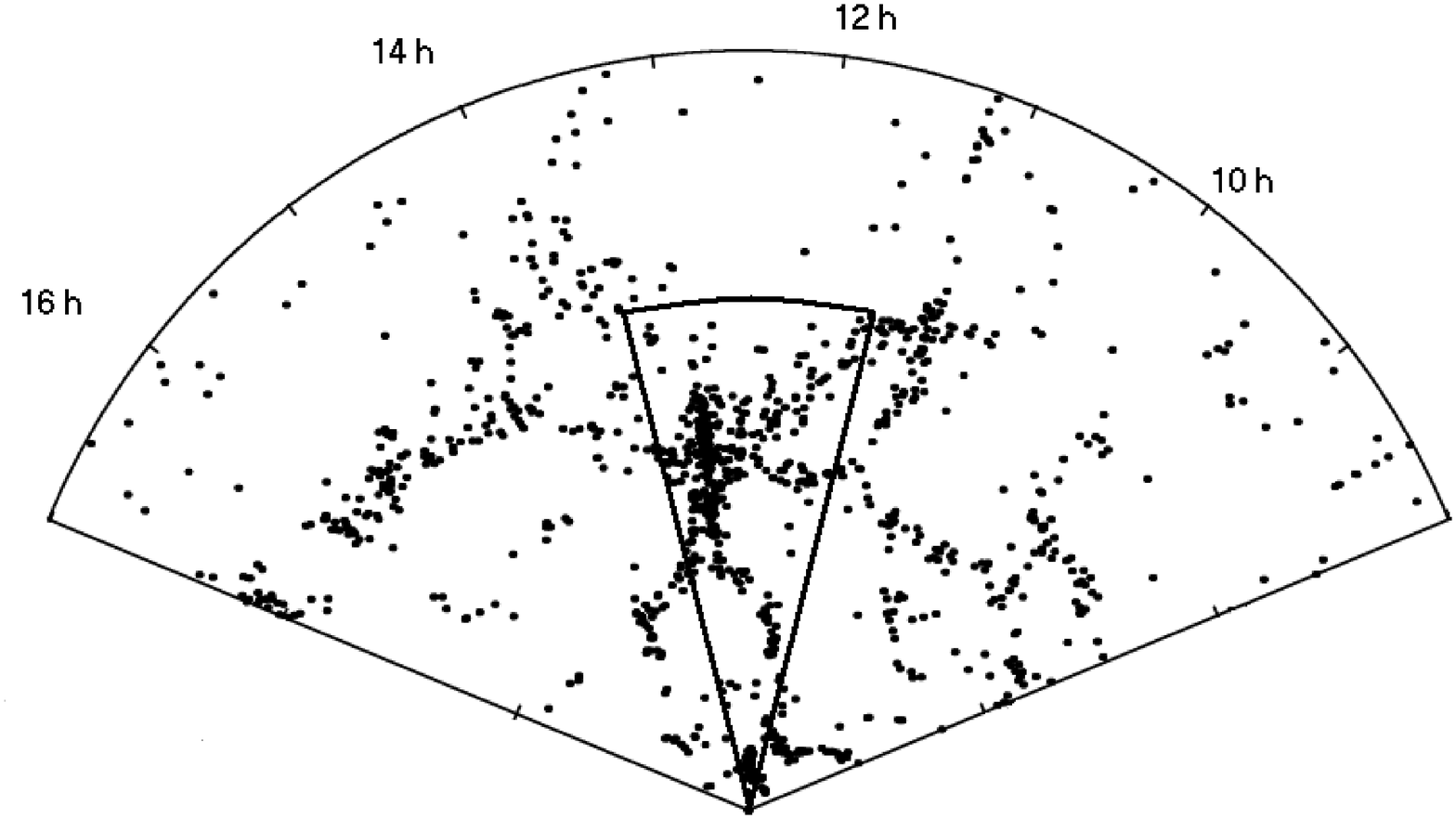} \caption{Left: The
original redshift data from de Lapparrent et al. (1986) is shown in this
plot with redshift increasing from zero at the lower apex to 15,000 km/s
along the arc of right ascension.  Right: The
Gregory and Thompson (1978) redshift survey area is superimposed on
a re-plot of the de Lapparrent et al. (1986) data.  Because the core of A1367 is
at Dec. = 21$^o$, it is not included in this slice, and its Finger
of God feature is not present.\label{fig4}}
\end{figure*}

\section{The Second Wave Redshift Survey Work}
Other research groups began to substantiate the new void and
supercluster structural features in the galaxy distribution. The
field grew so rapidly in the mid-1980's -- a time when CCD cameras
began to replace image intensifiers -- that it is virtually
impossible to make the last half of Table 1 complete.  One important
group that contributed to the second-wave included Robert Kirshner,
Augustus Oemler,Jr., Paul Schechter, and Stephen Shectman. During
the late 1970's they measured galaxy redshifts in three small survey
fields, each separated by 35$^o$ on the sky (in the direction of the
constellations Bootes and Corona Borealis) with the aim of
determining the galaxy luminosity function.  After learning about
voids from the early redshift survey papers, they noticed a
deficiency of galaxies in the redshift interval between 12,000 km/s
and 18,000 km/s in all three of their narrow survey fields. On this
basis they speculated that the entire region between their three
small survey fields could be ``A million cubic megaparsec void''
(Kirshner 1981). In the end (Kirshner et al. 1987), the volume of
the Bootes void was remeasured to be 1/3 the value quoted in the
title of their 1981 paper.  Its radius is $\sim$34h$^-$$^1$ Mpc.

Martha Haynes, Ricardo Giovanelli and their friend Guido Chincarini began
redshift surveys in 1982 based on observations of the 21 cm emission line
of neutral hydrogen (Chincarini et al. 1983).  Giovanelli and Haynes continued
in subsequent years to make many seminal contributions to our knowledge of
the neutral hydrogen content of galaxies and to the structure and nature of
superclusters.  One of the better examples of their redshift survey work is
paper V in their series of eight papers on the Pisces-Perseus supercluster
(Wegner, Haynes and Giovanelli 1993).

The CfA galaxy redshift survey reached
full stride by the mid-1980's when this group began to push deeper than m = 15.
The first dramatic CfA results came with the publication of de Lapparent,
Geller and Huchra (1986), a wide angle survey to m = 15.5.
Figure 4a is a plot of the data used by de Lapparent et al. 1986.  For comparison sake,
Figure 4b is the same cone diagram with an overlay added showing the
Gregory and Thompson (1978) survey area.

\section{Summary Statement on the Early Redshift Survey Work}
IAU Symposium No. 124 entitled ``Observational Cosmology'' was held
in Beijing, August 25-30, 1986.  Allan Sandage gave the invited
opening presentation.  In his presentation,
Sandage (1987) says the following about the early redshift survey work:

Gregory and Thompson (1978) ``marks the discovery of voids, which
have become central to the subject [of the large scale structure].  Prior
work by Einasto et al. (1980 with earlier references), Tifft and Gregory (1976),
and Chincarini and Rood (1976) foreshadowed the development, but the
Gregory and Thompson discovery is generally recognized as the most
convincing early demonstration.  Rapid developments in the mapping of
various filaments and voids include the studies of Tarenghi et al. (1979),
Gregory et al. (1981), Kirshner et al. (1981), Gregory and Thompson (1984),
Chincarini et al. (1983), and Huchra et al. (1983).  A general review is given
by Oort (1983).''

IAU Symposium No. 124 was held immediately after the the March 1,
1986, publication of the paper by de Lapparent, Geller, and Huchra
(1986) showing the ``Coma stickman'' in the center of the CfA cone
diagram (Fig. 4).


%% Included in this acknowledgments section are examples of the
%% AASTeX hypertext markup commands. Use \url without the optional [HREF]
%% argument when you want to print the url directly in the text. Otherwise,
%% use either \url or \anchor, with the HREF as the first argument and the
%% text to be printed in the second.

\section*{Acknowledgments}

This research has made use of NASA's Astrophysics Data System
Bibliographic Services.

\begin{deluxetable}{p{4.5cm}rlp{5.3cm}} \tablecolumns{4} \tablewidth{0pc} \tablecaption{Redshift
Survey Timeline} \tablehead{ \colhead{Authors} & \colhead{Citations}
 & \colhead{3D Plot} & \colhead{Description}} \startdata
Chincarini \& Rood 1971 & 107 & \nodata & Perseus cluster virial analysis \\
Chincarini \& Rood 1972a & 78 & \nodata & Perseus, Cancer, Coma, A2197/9 \\
Chincarini \& Rood 1972b & 46 & \nodata & Coma, A2197/9, NGC4272 \\
Tifft \& Gregory 1973 & 21 & \nodata & Coma redshift data only \\
Tifft, Jewsbury, Sargent 1973 & 16 & \nodata & Cancer cluster virial analysis \\
Chincarini \& Martins 1975 & 18 & \nodata & Seyfert Sextet redshifts \\
Tifft \& Tarenghi 1975a & 25 & \nodata & A1367 velocity dispersion \\
Tifft \& Tarenghi 1975b & 30 & \nodata & Coma radio galaxy redshifts \\
Tifft, Hilsman \& Corrado 1975 & 19 & \nodata & NGC507 cluster virial analysis \\
Chincarini \& Rood 1975 & 42 & radial & Survey 14.2$^o$ W of Coma. \\
Tifft \& Gregory 1976 & 135 & cone & Coma r=6$^o$ "devoid" foreground \\
de Vaucouleurs et al. 1976 & \nodata & \nodata & 2nd RCBG \\
Chincarini \& Rood 1976a & 87 & radial & Survey 14.2$^o$ W of Coma. Noticed a "segregation of redshifts". \\
Chincarini \& Rood 1976b & 29 & radial & Pegasus I \& II redshift survey \\
Tifft \& Tarenghi 1977 & 14 & \nodata & Redshifts of Coma radio galaxies \\
Davis, Geller \& Huchra 1978 & 82 & \nodata & m=13.0 mean mass density only \\
Gregory \& Thompson 1978 & 305 & cone & Coma/A1367 11$^o$x23$^o$ to m=15.0 Discovered voids with r$> $20h$^-$$^1$ Mpc \\
Joeveer, Einasto \& Tago 1978 & 114 & cone & Found "large holes" in 70$^o$x70$^o$ area \\
Tarenghi, Tifft, Chincarini, Rood \& Thompson 1979 & 127 & cone & Hercules redshift survey with void in foreground \\
Tarenghi, Chinicarini, Rood \& Thompson 1980 & 90 & cone & Hercules supercluster analysis \\
Gregory, Thompson \& Tifft 1981 & 173 & cone & Perseus supercluster with large voids \\
Chincarini, Thompson \& Rood 1981 & 57 & cone & 44h$^-$$^1$ Mpc
filament Hercules to A2197/9 \\
 \cutinhead{Selected Second Wave Papers}
Kirshner, Oemler, Schechter \& Shectman 1981 & 413 & \nodata & 35$^o$x35$^o$ sparsely sampled to m=16.3.  Bootes void r=34h$^-$$^1$ Mpc \\
Davis, Huchra, Latham \& Tonry 1982 & 241 & cone & m=14.5 north of Dec=0$^o$ \\
Einasto, Joeveer \& Saar 1980 & 166 & \nodata & Overview of large scale structure \\
Chincarini, Giovanelli \& Haynes 1983 & 39 & \nodata & HI redshifts Coma/A1367 bridge \\
de Lapparent, Geller \& Huchra 1986 & 772 & cone & 6$^o$x117$^o$ survey to m=15.5 \\
Wegner, Haynes \& Giovanelli 1993 & 69 & cone & Pisces-Perseus supercluster \\
\enddata
\end{deluxetable}

%% \end{document}

%%
%% End of file `table.tex'.

%% \endinput
%%
%% End of file `table.tex'.

%% This example uses \plotone to include an EPS file scaled to
%% 80% of its natural size with \epsscale. Its caption
%% has been written to indicate that additional figure parts will be
%% available in the electronic journal.

%% Here we use \plottwo to present two versions of the same figure,
%% one in black and white for print the other in RGB color
%% for online presentation. Note that the caption indicates
%% that a color version of the figure will be available online.
%%

%\begin{figure}
%\plottwo{f2.eps}{f2_color.eps} \caption{\label{fig2}}
%\end{figure}

%% The following command ends your manuscript. LaTeX will ignore any text

\end{document}